\documentclass[prb,twocolumn,tightenlines,floatfix,showpacs]{revtex4}
\usepackage{graphicx}
\newcommand{\beq}{\begin{equation}}
\newcommand{\eeq}{\end{equation}}
\newcommand{\bea}{\begin{eqnarray}}
\newcommand{\eea}{\end{eqnarray}}

\begin{document}

\title{Kinetics of macroion coagulation induced by
multivalent counterions}

\author{T. T. Nguyen and B. I. Shklovskii}

\affiliation{Theoretical Physics Institute, University of Minnesota, 116
  Church St. Southeast, Minneapolis, Minnesota 55455}

\begin{abstract}
Due to the strong correlations between multivalent
counterions condensed on a
macroion, the net macroion charge changes sign
at some critical counterion concentration. 
This effect is known as the charge inversion. Near
this critical concentration the macroion net charge is small.
Therefore, short range attractive forces
between macroions dominate Coulomb repulsion 
and lead to their coagulation. The kinetics
of macroion coagulation in this
range of counterion concentrations is studied. 
We calculate the
Coulomb barrier between two approaching 
like charged macroions
at a given counterion concentration. Two different
macroion shapes (spherical and rod-like) are considered.
A new ``self-regulated" regime
of coagulation is found. 
As the size of aggregates increases, their charge and
Coulomb barrier also grow and diminish the
sticking probability of aggregates.
This leads to
a slow, logarithmic increase of the aggregate size with time.
\end{abstract}


\maketitle

\section{Introduction}

Water solutions of strongly charged particles (macroions)
with  multivalent counterions with large charge $Z$
($Z$-ions) are important in
physics, chemistry, biology, chemical engineering
and environmental science.
Colloidal particles, charged membranes, double helix DNA,  
actin and other polyelectrolytes are 
examples of different macroions. 
Multivalent metallic ions, dendrimers, charged micelles,
short DNA helices or other short 
PE can play the role of $Z$-ions.

We concentrate here on strongly asymmetric solutions in 
which size and charge of macroions are much larger 
than those of $Z$-ions. As the simplest example, we  
consider macroions as
negatively charged rigid spheres with charge $-Q$ and 
radius $R$ in solution with compact positive $Z$-ions 
with the size $a \ll R$ and charge $1 \ll Z \ll Q$. This can be a 
solution of positive
latex particles with very short DNA helices\cite{Grant,Gotting}
or hematite particles with polyacrylic acid\cite{Buffle}.
 
In such solutions, each sphere adsorbs many $Z$-ions.
They strongly repel each other at the surface of the sphere
and form a strongly correlated two-dimensional 
liquid reminding Wigner crystal.
When a new $Z$-ion approaches this liquid, 
it repels nearest $Z$-ions, creates 
a correlation hole or an oppositely charged image,
which provides 
attraction of $Z$-ion to the surface\cite{Perel99}
in addition to what mean field theories predict.
Therefore, when concentration of $Z$-ions, $c$, 
reaches some critical value $c_0(s)$, 
(which depends on concentration of 
macroions particles $s$) the net charge of each macroion, $Q^*$,
(which includes all adsorbed $Z$-ions) flips its sign\cite{Nguyen}.
On Fig. 1 the ``neutrality" line $c_0(s)$ is shown
in plane ($s$, $c$) together with 
two signs of $Q^*$ which it separates.
\begin{figure}
	\resizebox{6.5cm}{!}{\includegraphics{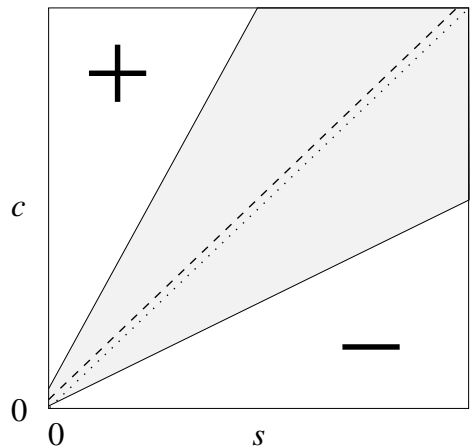}}
    \caption{Phase diagram of a solution of 
    negative spheres with compact $Z$-ions in
    plane of their concentrations ($s$,$c$). The dotted line corresponds     
	to the isoelectric composition $c=sN_i$,
	where $N_i$ is the number of $Z$-ions needed to neutralize one sphere.
	The dashed line corresponds to
      the concentration of $Z$-ions, $c = c_{0}(s)$,
      where net charge of a sphere with adsorbed $Z$-ions
      crosses zero. The two solid lines define the external 
      boundary of domain where spheres coagulate. 
      Plus and minus are the signs of the net charge
      of free spheres above and below the dashed curve.}
    \label{fig:phase}
\end{figure}

Correlations between $Z$-ions on the sphere surface
also lead to another
interesting effect. The correlation energy
per $Z$-ion is lower
at higher $Z$-ion concentration. This means
when two spheres touch each other, at the place
of contact where the $Z$-ion concentration doubles,
correlation energy is gained\cite{Rouzina}. 
This energy gain leads
to an attraction between spheres which
is stronger than van der Waals forces and
makes possible 
their coagulation.
Coagulation is a key part of many industrial processes
such as paper production, extraction of 
minerals, proteins and other macroions from solutions,
or treatment of waste waters. 
On the other hand, there are many other cases where coagulation 
should be avoided. Delivery
of short modified DNA molecules DNA adsorbed
on the surface of positively 
charged latex particles is a good example\cite{Grant,Gotting}.

Because colloidal solutions are stable due to 
the Coulomb repulsion between
particles, coagulation is usually achieved by a large 
concentration of monovalent salt, which screens out these charges.
When the concentration of salt grows beyond the coagulation
threshold nothing new happens and macroions stay coagulated.
If, instead of monovalent salt, we deal with $Z$-ions (add Z:1 salt)
then charge inversion changes this situation. Coagulation happens only in
the range of concentrations, which is close to 
the neutrality line 
(shown by gray in Fig. 1), so that 
the Coulomb repulsion is so weak that 
the free energy is gained when spheres touch each other.
Thus when, say, the concentration of $Z$-ions, $c$,
grows at fixed $s$, coagulation happens 
at some critical concentration, $c_c$, 
while at a larger concentration, $c_d$, aggregates of spheres
dissolve, because spheres acquire large positive net charge.
In our previous paper\cite{Nguyen}, we derived 
expressions for $c_c(s)$ and $c_d(s)$. These
curves are plotted on 
Fig. \ref{fig:phase}. 
Remarkably, phase diagrams of this type were discovered 
experimentally more than half a century ago for complexes of 
strongly
oppositely charged proteins\cite{deJong}.
However, in Ref. \onlinecite{deJong} only the charge of large
aggregates of macroions (coacervate droplets) 
was measured and the net
charge of a single macroion was not discussed.
No explanation was proposed for charge inversion of 
aggregates. In Ref. \onlinecite{Nguyen} we suggested 
an equilibrium theory of the phase diagram of Fig. 
\ref{fig:phase} and showed that
aggregates of spheres and isolated spheres change sign of their
net charge at the same line $c_0(s)$.

In this paper we go beyond equilibrium 
statistical physics and study kinetics 
for different domains of the phase diagram.
There are two main slow processes in the problem:
charge inversion and coagulation. 
The first one is slow in the upper left area of the phase diagram 
when free macroions are strongly overcharged. In this case,
the growth rate of the macroion positive charge 
is limited by a large repulsive 
Coulomb barrier for new $Z$-ions.
Activation above this barrier is necessary
for a $Z$-ion to come close enough 
to macroion in order to feel attraction to its image in 
the strongly correlated liquid (SCL) of already adsorbed 
$Z$-ions. 

In the coagulation domain surrounding
the neutrality line charges of macroions are relatively small 
so that charge inversion is a fast process. 
On the other hand, 
coagulation of macroions can be very slow and take hours.
Suppose we mix latex spheres
with $Z$-ions in such concentrations that the corresponding 
point is in the gray area of the phase diagram and watch
how the mass of aggregates grows as a function of time.
Although the net charge $Q^*$ in the gray area 
is smaller than the bare charge $Q$, its absolute value
can be much larger than $Z$ and the Coulomb barrier of repulsion
of two macroions can be much larger than
the barrier for charge inversion, which is
created by the Coulomb interaction of a $Z$-ion with 
the macroion net charge $Q^*$. Therefore, in this
paper, we assume $Z$-ions are always in equilibrium
and study the kinetics of macroion coagulation only.

We first concentrate on the short time kinetics
in which doublets
of macroions appear. We calculate the Coulomb barrier
between two approaching macroions as a function 
of the distance from the neutrality line.
We measure this distance by the variable 
\begin{equation}
U = \frac{k_BT}{Ze}\ln\left[1+\frac{c-c_0(s)}{c_0}\right]~,
\label{voltage}
\end{equation}
where the critical concentration
\beq
c_0(s)=c_0+sN_i.
\eeq
Here $N_i$ is the number of $Z$-ions needed to neutralize one sphere.
The concentration $c_0$ is the concentration of free
$Z$-ions which is in equilibrium
with neutralized spheres. 
It is equal to\cite{Perel99}
$c_{s}\exp(\mu_{SCL}/k_BT)$, where
$c_{s}$ is the
three dimensional concentration of the surface layer of the SCL of
$Z$-ions and $\mu_{SCL}$ is the chemical potential of $Z$-ions
in SCL. It is negative and $|\mu_{SCL}| \gg k_BT$ so that  $c_0$ is very
small.

In Ref. \onlinecite{Nguyen}, the net charge of the sphere
was shown to be proportional to $U$ near the neutrality line:
\beq
Q^*=CU~,
\label{eq:QCU}
\eeq
where $C$ is the 
capacitance of the macroion coated by the "metallic film"
of SCL of $Z$-ions.
For a sphere $C = \varepsilon R$, where  
$\varepsilon$ is dielectric constant of water. Because $U$ depends
on the correlation chemical potential $\mu_{SCL}$ of $Z$-ions
in the SCL, we call it the correlation voltage.
From Eq. (\ref{voltage}), one sees that
if $c > c_0(s)$ spheres
are overcharged (positive), while in the opposite case  $c >
c_0(s)$ they are undercharged (negative).
Detail derivation of Eq. (\ref{eq:QCU}) can be
found in Sec. II of Ref. \onlinecite{Nguyen} (See Eq. (26) of
this reference).
 
Our main result is that the height of the
Coulomb barrier between two spheres is
\begin{equation}
V_{max} = -\alpha CU^{2}~.
\label{eq:vmaxintro}
\end{equation}
where $\alpha$ is a numerical factor 
of the order of unity (at week screening, $\alpha=0.3$,
at very strong screening, $\alpha=1$).
%
%
%
Thus the rate $\nu$ at which doublets appear has the form
\beq
\nu(U)=\nu_0\exp(-\alpha CU^2/k_BT)~.
\label{rate}
\eeq
Exponential factor of Eq. (\ref{rate}) 
is, of course, the probability
of activation above the Coulomb barrier between two macroions.

Eq. (\ref{eq:vmaxintro}) suggests that the Coulomb
barrier between aggregates increases as their size grows.
This leads to the decreasing sticking probability
between two aggregates as their size increases.
This, in turn, leads to the slowing down of
the growth of the aggregate size with time. We call this
the ``self-regulated" aggregation and show that
in this case
an aggregate size
increases only as a logarithmic function of time
\beq
{\cal R}(t)\simeq{\cal R}_1\ln(t/t_1),
\label{eq:self}
\eeq
where the size and time constant ${\cal R}_1$ and $t_1$
will be given in Sec. III.
To the best of our knowledge, this logarithmic 
behaviour is
reported for the first time.

However, ``self-regulated" aggregation cannot continue 
forever.
When aggregates are so large that their time of activation
above the Coulomb barrier is longer than the
time for one macroion to desorb from an aggregate,
one enters a Lifshitz-Slezov (LS) regime of coagulation
where the aggregate size increases linearly
with time with a very long time constant.
In this regime, aggregates gains size by adsorbing free
spheres which desorb from other aggregates.

Screening can diminish the Coulomb barrier substantially.
At strong screening, the LS regime may never
be reached. Instead the height of Coulomb barrier saturates
when the aggregate size reaches the screening length $r_s$.
In this case, the ``self-regulated" aggregation is 
followed by the
reaction limited aggregation 
where Coulomb barrier is constant for any aggregates
with size greater than $r_s$ and the
average aggregate size increases roughly quadratically in time.

There have been various experiments studying the kinetic
of macroion coagulation induced by $Z$-ions, such as
latex particles complexed with
short DNA segments\cite{Grant} or hematite particles
complexed with polyacrylic acid\cite{Buffle}.
In both experiments, the author observed
an exponential increase in the rate of coagulation when
the neutrality line is approached. At the same time,
as the $Z$-ions
concentration increases across this line, the 
electrophoretic mobility
of the macroions changes sign suggesting a charge 
inversion effect. When $Z$-ions concentration are much larger or
smaller than the neutrality composition, macroions are
under- or overcharged and the Coulomb barrier
exponentially diminishes their sticking probability.
These results are in qualitative agreement
with Eq. (\ref{rate}). 
However, instead of $\ln[\nu(U)]\propto U^2$, 
the authors of Ref. \onlinecite{Buffle}
observed a plateau in the aggregation rate around the
neutrality point $U=0$.
This may be due to the limited time resolution
of the experiment at the initial fast stage
of coagulation.

The paper is organized as following. In Secs. II and IV, we study
the kinetic barrier for two approaching macroion with spherical
shape and rod-like shape at a given $Z$-ion concentration.
In section III, we discuss different stages
of macroion coagulation, namely how the coagulation rate
crossovers from a diffusion limited regime to the new
``self-regulated" regime
and finally to LS regime. In the conclusion,
we summarize our results.

\section{Coulomb potential barrier between two approaching spheres}

Let us start by calculating the Coulomb barrier between two 
approaching spheres. 
Because the charge of each sphere is not fixed but
self-adjusts (by releasing or absorbing $Z$-ions)
according to their positions,
one has to calculate self-consistently the Coulomb repulsion
between spheres and their charges at a given
separation.
The capacitor charging picture of Eq. (\ref{eq:QCU}) 
offers a very convenient way of 
doing this. Indeed,
the voltage $U$ depends only on the concentration of bulk $Z$-ions and
the surface charge density of the spheres (through the chemical
potential $\mu_{SCL}$)
and, therefore, is constant
for a given $c$ and sphere surface charge density 
$\sigma$. Because the SCL of $Z$-ions on the macroion
surface behaves as a metal, 
one can view the
system of spheres and their aggregates as a system of conductors
under a constant charging potential $U$. 
Thus, one can calculate the net
charge of not only spheres but also
their aggregate of any size
by using the appropriate capacitance $C_{\rm aggregate}$ 
instead of $C$. In the
same way, the kinetic Coulomb potential barrier between any two 
aggregates can also be calculated. 

Because, $Z$-ions are much more mobile than 
spheres, the process of charging up the
spheres is much faster than coagulation
and at any instance during the coagulation
process, the $Z$-ions distribution is
in equilibrium.
With this assumption,
one can write the Coulomb interaction energy between two spheres
when they are at distance $r$ from each other as:
\beq
V(r)=\frac{C_{11}(r)+2C_{12}(r)+C_{22}(r)}{2}U^2- \varepsilon RU^2~,
\label{eq:barier}
\eeq
where $C_{11}(r)$, $C_{22}(r)$ and $C_{12}(r)$ are, respectively,
the self capacitances of the spheres 1 and 2 and their mutual capacitance.
The capacitance of an isolated sphere of 
is $\varepsilon R$.

When dealing with a system under constant charging potential,
the free energy $F(r)$ 
of the system must include the work of the source
(in our case, the population of free $Z$-ions) 
to maintain this potential
\beq
F(r)=V(r)-\sum_{i,j=1}^{2}C_{ij}(r)U^2=-V(r).
\eeq
%
It is the lowering in the free energy compared to the
reference system of two completely neutralized spheres with
the rest of $Z$-ions free.

The capacitance $C_{11}$, $C_{22}$ and $C_{12}$ has been calculated for
two spherical conductors\cite{Smythe}:
\beq
C_{11}(r)=C_{22}(r)=\varepsilon R\sinh\beta
		\sum_{n=1}^{\infty}\sinh^{-1}[(2n-1)\beta],
\label{eq:Cself}
\eeq
\beq
C_{12}(r)=-\varepsilon R\sinh\beta
		\sum_{n=1}^{\infty}\sinh^{-1}(2n\beta),
\label{eq:Cmutual}
\eeq
where $\beta >0$ satisfies $\cosh\beta=r/2R$.

\begin{figure}
	\resizebox{7cm}{!}{\includegraphics{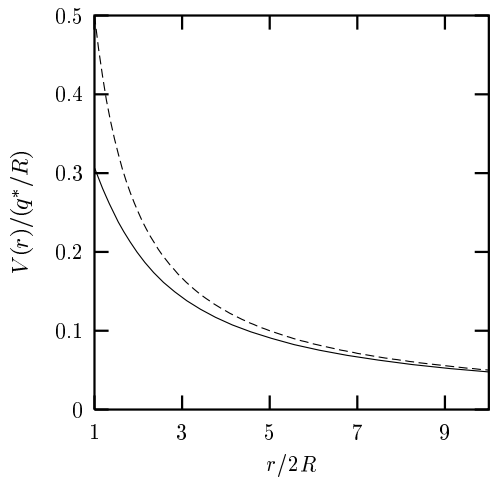}}
    \caption{The Coulomb potential barrier between two
	spheres (the solid line). For comparison, the Coulomb potential
	barrier for two spheres with fixed charge is plotted (the
	dashed line). Note that, although not plotted,
	at distances very close to $r/2R=1$, the interaction energy
	drops below zero due to the short range correlation attraction.}
\label{fig:barier}	
\end{figure}

Substituting Eqs. (\ref{eq:Cself}) and (\ref{eq:Cmutual}) into 
Eq. (\ref{eq:barier}), one can easily find the cost in the free energy
$-V(r)$ in moving two spheres from infinity to the distance $r$.
The result is plotted by the solid line in Fig. \ref{fig:barier}. 
For comparison, the potential barrier for the case when the spheres keep
their charge $q^*=C(\infty)U=\varepsilon RU$ fixed when 
approaching each others is also plotted.
The maximum of the potential barrier is located very close
to the distance $r=2R$ where two spheres touch
each other and is equal to
\beq
V_{max}\simeq 0.3\varepsilon RU^2.
\label{eq:vmax1}
\eeq
As one can see, at the maximum of the potential barrier, 
the self adjustment of the sphere charge reduces
this barrier height by about 40\% compared to the barrier
of $0.5\varepsilon RU^2$ one would get if the sphere charges were
kept fixed upon approaching.

Due to this Coulomb barrier, the expression 
for the rate at which spheres come 
to each other to form doublets contains an exponential
factor related to the probability of activation above
this barrier as shown in Eq. (\ref{rate}).

In the above calculation, it is assumed the Coulomb potential is 
unscreened. In reality, there is always a finite concentration of 
monovalent salt in water solution, which leads
to the screening of Coulomb interaction at distance larger than
the Debye-H\"{u}ckel screening length $r_s$. If $r_s$ is larger
than the averaged distance between $Z$-ions at the surface of
an aggregate, the correlation between $Z$-ions remains 
unscreened. Therefore, the concentration $c_0$, and correspondingly,
the charging potential $U$ remain constant. Thus, in this
regime, screening influences the process of coagulation only
through the change in the capacitance  of each sphere.

When the radius of the aggregate is smaller than the screening
radius, $R < r_s$, the Coulomb interaction is not screened
when the aggregates touch each other, therefore
the above result remains valid. In the opposite case, when
$R > r_s$, the capacitance of each isolated aggregate is
that of a plane capacitor with thickness $r_s$ and area 
$4\pi R^2$,
$C_{1,2}(\infty)=\varepsilon R^2/r_s$. Using this planar capacitor
approximation, it is not difficult to estimate the change in 
the total capacitance of the system when the two aggregates
touch each other. Indeed, simple geometric calculation
gives the reduction in the area of the planar
capacitor when the two spherical aggregates touch each other
is $4\pi Rr_s$. Thus the change in the capacitance
of the system is
$\Delta C=C_{11}+2C_{12}+C_{22}-C_1(\infty)-C_2(\infty)=\varepsilon R$.
Thus, in this case the maximum of the potential barrier is
\beq
V_{max}= \varepsilon RU^2~, ~~~~~~ (R >> r_s).
\label{eq:vmaxrs}
\eeq

Comparing this results with Eq. (\ref{eq:vmax1}), one sees that,
beside a numerical factor, screening does not affect the 
height of the potential barrier between two approaching
spheres as long as $r_s$ is larger
than the average distance between $Z$-ions on the spheres.

\section{Later stages of coagulation}

In previous sections, we discussed the kinetic barrier when two macroions
approach each other to form a doublet. This fast process happens at the
initial stage of coagulation where the concentration of
free macroions is large. Doublets coagulate with free macroions
or other doublets to form bigger macroions aggregates. 
This section deals with later stages of coagulation.

The problem of coagulation is generally complicated and can be solved
exactly only in special cases. However, the main physics can be
captured if one works with the dominant
aggregate size ${\cal R}(t)$ (with concentration ${N}(t)$) at a given time
assuming this typical aggregates carry all the mass of the macroions.

Let us start from the initial stage of coagulation when 
the aggregate size are small, so that at small enough $U$ (close
to the neutrality line) the Coulomb
barrier between aggregates is smaller than $k_BT$.
In this case, the Coulomb barrier of Eq. (\ref{eq:vmax1}) 
has little effect
and the probability
of sticking for two aggregates 
is of the order of unity. This is the regime of the
well-known diffusion limited
aggregation. 
The concentration of the typical size aggregates decreases as inverse
of time\cite{colloid}:
\beq
{N}(t)=\frac{s}{1+t/\tau_{\rm diff}} 
\simeq s \frac{\tau_{\rm diff}}{t}~~~ {\rm for }~~ t \gg \tau_{\rm diff},
\eeq
where $\tau_{\rm diff}\sim \eta/k_BT s$, $s$ is the
initial concentration of free macroions and $\eta$ is the viscosity 
of water. Thus, the rate
of coagulation in this early stage is relatively constant independent
of $U$
and equal to the rate of coagulation at the neutrality line
where the charging potential $U$ is exactly zero and
there is no Coulomb barrier.

The regime of diffusion limited aggregation stops when
the typical size of the macroion aggregates reaches such a size
that the Coulomb barrier is larger than $k_BT$.
Using Eq. (\ref{eq:vmax1}), one sees that
this regime is reached when the typical size equal
\beq
{\cal R}_1\simeq k_BT/\varepsilon U^2.
\eeq
Assuming an aggregate has a fractal dimension $d_f$ 
(in practical situation, $d_f\simeq 2$) so that the number
of spheres in it is $({\cal R}/R)^{d_f}$,
the time at which ${\cal R}_1$ is reached is
\beq
t_1=\tau_{\rm diff}(k_BT/\varepsilon U^2R)^{d_f}.
\eeq

When $t> t_1$, the Coulomb
barrier $V_{max}\simeq \varepsilon {\cal R}U^2$ 
has to be taken into account. In this regime, the dominant
contribution to the rate of coagulation comes from
the tunneling of aggregates through the Coulomb barrier.
The change in the concentration of typical aggregates
obeys the equation\cite{colloid}:
\beq
\frac{d{N}(t)}{dt}=-\kappa{D}{\cal R}^2{N}^2
	\exp(-V_{max}/k_BT)
\label{eq:mastereq2}
\eeq
where the numerical factor has been dropped on the right hand
side and the factor
$\kappa$ satisfies $\kappa^2=-(d^2 V(r)/d r^2)_{max}/2k_BT$.
Since the typical width of the Coulomb barrier is $\cal R$,
$\kappa \simeq \sqrt{V_{max}/{\cal R}^2k_BT}$.
Using the Stoke formula,
$D\simeq k_BT/\eta {\cal R}$,
one can solve
Eq. (\ref{eq:mastereq2}) 
assuming that the typical
size aggregates consume all the mass of the macroions:
\beq
{\cal R}={\cal R}_1({N}_1/{N})^{1/d_f},
\label{eq:fractal}
\eeq
where ${N}_1={N}(t_1)$. 
The concentration of typical
size aggregates in this regime decreases logarithmically with time:
\beq
{N}(t)\simeq {N}_1\left[
				\ln	\left( \frac{t-t_1}{t_1d_f}+e \right)
		\right]^{-d_f}
\eeq
where ${N}_1={N}(t_1)$. Eq. (\ref{eq:self}) for
the size of the typical aggregates can be obtained easily
using the relationship (\ref{eq:fractal}).

As of our knowledge, this slow logarithmic kinetics was
never reported in literature. It is the result of the increase
in Coulomb barrier when aggregate size grows. This diminishes
their sticking probability, which in turn slows down the kinetics
from the linear size increase to a logarithmic one. 
We thus call this regime ``self-regulated" aggregation.

In the regimes we considered so far, the increase in the aggregate
mass is caused 
by a collision between two aggregates of size ${\cal R}(t)$.
When the typical aggregate size becomes very large, the Coulomb
barrier between two approaching aggregates becomes so high
that the corresponding coagulation rate (which is proportional
to $\exp(-V_{max}/k_BT)$) becomes very small, a different and
faster aggregate growth mechanism, namely the Lifshitz-Slezov
(LS) one,
comes into play. In this mechanism, the large aggregates do not 
collide into each other and the primary mechanism for
aggregate growth is no longer due to real space diffusion.
Instead, the change in their size
is provided by the releasing and adsorbing of  spheres
which have much smaller kinetic barriers. This mechanism
provides a faster coagulation process because
the binding energy of a sphere in a very
large aggregate and the Coulomb barrier between
a very large aggregate and a sphere are finite
and do not depend on the aggregate size.
One can think of LS mechanism
as a diffusion in size space\cite{LS}.
Interestingly, the concentration of typical
size aggregates in this mechanism also decreases
as inverse of time
\beq
{N}(t) \propto \tau_{\rm LS}/t,
\eeq
where the time constant $\tau_{\rm LS}$ is much longer than
$\tau_{\rm diff}$ because it is proportional to
the concentration of free macroions in solution 
%
\beq
\tau_{\rm LS} \propto \exp(-E/k_BT)
\eeq
Here $E$ is 
the binding energy of a sphere in an aggregate which
does not depend on the aggregate size.

One can easily find the size of the typical aggregate 
at which LS mechanism is important
by comparing the detaching time $\tau_{LS}$ of
a small macroion from the aggregates with
the activation time for two aggregates to go through
the Coulomb barrier which is of the order
$\exp(-\varepsilon {\cal R}U^2/k_BT)$. This gives
\beq
\varepsilon {\cal R}U^2=E.
\eeq
Thus the typical size at which LS regime starts is:
\beq
{\cal R}_2=E/\varepsilon U^2 .
\eeq

\begin{figure}
	\resizebox{7.5cm}{!}{\includegraphics{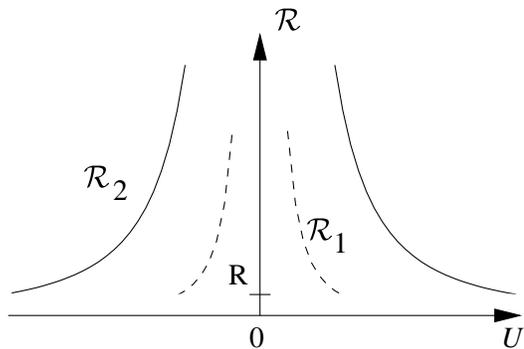}}
    \caption{The typical aggregate sizes ${\cal R}_1$ and ${\cal R}_2$
	at which aggregate growth mechanism crosses over 
	from diffusion limited
	regime to ``self-regulated" regime and then to Lifshitz-Slezov
	regime. 
	}
\label{fig:R}	
\end{figure}

A phase diagram of 
different regimes of coagulation is shown in
Fig. \ref{fig:R}. With increasing time, $\cal R$ grows
along a vertical line starting from ${\cal R}=R$.
Below the dotted line which represents
${\cal R}_1(U)$, the coagulation is diffusion limited and
the aggregate mass increases with time linearly as $t/\tau_{\rm diff}$.
Above the solid line which represents ${\cal R}_2$,
the coagulation process is of LS nature and the aggregate
mass also increases linearly with time $t/\tau_{\rm LS}$.
In between these two lines is regime of ``self-regulated" aggregation.
In this regime, the aggregate size increases
logarithmically.

Above, we dealt with
the unscreened Coulomb potential between two aggregates.
This is valid if the screening length of 
the solution is very large ($r_s > {\cal R}_2$).
For smaller screening length,
when the typical aggregate size reaches $r_s$, 
due to the fractal nature
of aggregates, the Coulomb barrier stops increasing\cite{footnote}.
In this case, LS regime cannot be reached. One goes from
the ``self regulated" regime to the regime of 
reaction limited aggregation
with constant Coulomb barrier and almost zero probability of 
macroion detaching from the aggregates.
One can calculate the change in the concentration $N$ in
this regime using the Eq. (\ref{eq:mastereq2}).

Denote by $t_s$ the time at which
the typical aggregate
size reaches $r_s$ (from Eq. (\ref{eq:self}),
$t_s \simeq t_1 \exp(\varepsilon r_s U^2/k_BT)$). 
At $t > t_s$,
the exponential factor in Eq. (\ref{eq:mastereq2}) becomes
constant, $\exp(-\varepsilon r_sU^2/kBT)$,
and so does the factor $\kappa$.
The solution of this equation shows 
a fast decrease in the
concentration of typical aggregates with time:
\beq
{N}(t) = {N}(t_s) \left[
				1+\frac{t-t_s}{\tau_{\rm react}}
				\frac{d_f-1}{d_f}
		\right]^{d_f/(1-d_f)} ,
\eeq
where $\tau_{\rm react}$ is an exponentially long time constant:
$$
\tau_{\rm react}=\frac{\eta}{k_BT{N}(t_s)}
		\frac{\exp(\varepsilon r_sU^2/k_BT)}
		{\sqrt{\varepsilon r_sU^2/k_BT}}~.
$$
In a typical situation, $d_f \simeq 2$, the number of spheres
in the typical
aggregates (which is inversely proportional to ${N}$)
increases as $t^2$.
Although this is a very fast kinetic,
it is still slower than 
an exponential growth of the cluster size 
suggested recently\cite{RLA,Witten}. 
We currently do not have a clear understanding of the
origin of
this theoretical result.

At even smaller screening length, $r_s < {\cal R}_1$, the Coulomb
barrier between two aggregates never becomes larger than $k_BT$
and one always stays in the regime of diffusion limited aggregation.

\begin{figure}
	\resizebox{6.5cm}{!}{\includegraphics{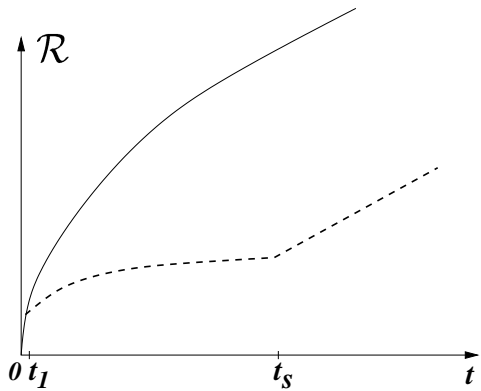}}
    \caption{Sketch of the dependence
	of the size $\cal R$ of typical aggregates as function of
	time $t$. The solid line corresponds to the
	case of very strong screening, $r_s < {\cal R}_1$,
	where Coulomb barrier is smaller than $k_BT$
	and aggregation is diffusion limited
	(${\cal R}\propto \sqrt{t/\tau_{\rm diff}}$). The
	dashed line corresponds to intermediate screening,
	${\cal R}_1 < r_s < {\cal R}_2$, where
	one goes from diffusion limited aggregation 
	to self-regulated aggregation (${\cal R}\propto \ln(t/t_1)$) 
	and to irreversible reaction limited
	aggregation (${\cal R}\propto t/\tau_{\rm react}$).
	}
    \label{fig:rt}
\end{figure}

The evolution of the typical aggregate size as a function
of time is plotted in Fig. \ref{fig:rt} for $r_s < {\cal R}_2$.

\section{Aggregations of rod-like polymers}
In this section, we would like to discuss the kinetic barrier
for  the coagulation process when the role of macroion
is played by a rigid polyelectrolyte (PE), such as DNA. We 
assume the length of PE molecules is smaller than its persistence
length so that each can be considered as rigid rod. Due to
the anisotropy of this problem, there is a number of different
paths of aggregation of rods.

\begin{figure}
	\resizebox{7cm}{!}{\includegraphics{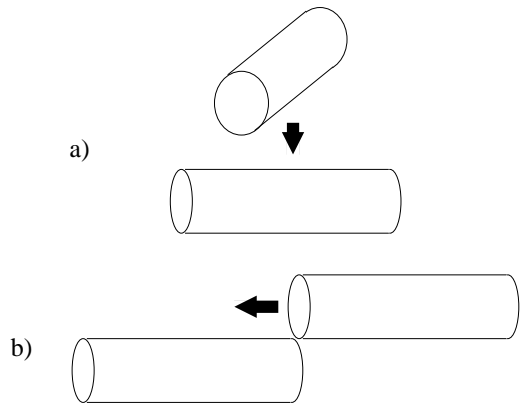}}
    \caption{Different paths of coagulation for two rigid rods.
	(a) The path studied in Ref. \onlinecite{Liu}. (b) 
	The path which according to our paper has lower kinetic
	barrier.}
    \label{fig:DNAkinetic}
\end{figure}

The authors of 
Ref. \onlinecite{Liu}
studied the kinetic barrier between two approaching rods
as a function of their orientation when the rods 
approach each other
in the direction perpendicular to their bodies
(Fig. \ref{fig:DNAkinetic}a). We believe
the kinetic barrier associated with this way of approaching
is too large
because of their large electrostatic
repulsion when placed side by side, especially when the screening
of solution is week.
In this section, we propose another path of approaching of two rigid
rods which has a lower kinetic barrier. Namely, the two rods 
approach in such a way that their 
centers of mass lie on a line parallel to
their longer axes (See Fig. \ref{fig:DNAkinetic}b). 
One can easily see that, in this way,
charges of the rods are kept further from each other 
than in Fig. \ref{fig:DNAkinetic}a. Thus
for the paths of Fig. \ref{fig:DNAkinetic}b,
the kinetic barrier between the rods is lower.

We can calculate the Coulomb barrier between two approaching rods
as in Sec. II, starting from the idea that both rods
are under constant charging voltage $U$.
The capacitance of a metallic rods of length $L$ and
radius $a$ is $C=\varepsilon L/2\ln(L/a)$ for
a weakly screening solution ($r_s >> L$) and 
is $C=\varepsilon L/2\ln(r_s/a)$ 
for stronger screening ($L >> r_s >> a$). Thus the
change in the capacitance of the system when the two
rods ends start touching other is 
$\Delta C \simeq \varepsilon L/\ln^2(L/a)$ 
for the weak screening case,
and is $\Delta C \simeq \varepsilon r_s/\ln^2(r_s/a)$
for stronger screening.

As the rods start to overlap, the short range attraction between
the rods starts to come into play, which partially compensate
the increases in Coulomb repulsion and reduced the kinetic barrier.
If the length of the overlap segment is $x$ and if
the attraction energy is $-\epsilon$ per unit length, 
the free energy of
the system can be written as 
$-\Delta C(x)U^2-x\epsilon$, which is 
$x(\varepsilon LU^2/\ln(L/a)-\epsilon)$
for the weak screening case and is 
$x(\varepsilon LU^2/\ln(r_s/a)-\epsilon)$
for the strong screening case.
Because, in the final stage when the two rods lie parallel to
each other, the attraction energy is assumed to win over their
Coulomb repulsion, $\epsilon > \varepsilon LU^2/\ln(r_s/a)$, one easily
sees that when the rods start to overlap, total energy start to decrease.
Thus the maximum of the potential barrier is at about the distance
at which the rods start to overlap.
\beq
V_{max}\simeq \left\{ \begin{array}{ll}
				\varepsilon LU^2/\ln^2(L/a) & {\rm for }~ r_s > L\\
				\varepsilon r_sU^2/\ln^2(r_s/a) & {\rm for }~ r_s < L.
				\end{array}
		\right.
\eeq
Obviously, this 
potential barrier is much smaller than that in the case the rods
approach each other in the direction perpendicular to their length.
In the latter case, the maximum of the potential barrier
is 
at the distant where they touch each other side by side
and
is equal to 
$\varepsilon LU^2/\ln(L/a)$ for $r_s > L$,
and 
$\varepsilon LU^2/\ln(r_s/a)$ for $r_s < L$.

At later stages of aggregation, collinear approaches should dominate as
well. This can explain why in many real and numerical
experiments, very elongated structures are seen.

\section{Conclusion}

In this paper, we provide a general framework
for calculation of
the Coulomb potential between two macroions
in solution of multivalent counterions and
study various stages in the coagulation of macroions.
The capacitance interpretation of Eq. (\ref{eq:QCU}) 
proves
to be very useful in calculating the net charge of
any macroions and aggregates of any shape. It also 
helps to easily calculate the Coulomb barrier between two
approaching macroions. Using this equation, we are able to
calculate and compare different paths of coagulation
for rod-like macroions.

We discuss several stages of coagulation in time.
In low salt, the coagulation process goes from
a diffusion limited regime to a ``self-regulated"
regime and finally to the regime Lifshitz-Slezov kinetic.
In the ``self-regulated" regime, as the aggregate size
increases, their Coulomb barrier increases diminishing their
sticking probability and slowing down the kinetic. As a result,
the aggregate size increases as a slow logarithmic function
of time instead of the standard linear relationship.
At higher salt concentration, Coulomb barrier is screened
and stop increasing after the aggregates reach a certain size
(of the order of the screening length $r_s$).
In this case, Lifshitz-Slezov regime cannot be reached. 
Instead, one reaches a reaction limited regime
where the Coulomb barrier is constant and
the aggregate size increases quadratically 
in time. At very high salt concentration, 
one even cannot reach the reaction limited aggregation
regime and always stays
in the regime of diffusion limited aggregations.


\begin{acknowledgments}
The authors are grateful to A. Yu. Grosberg and
S. Stoll for useful discussions. This  work is supported by 
NSF No. DMR-9985785.
\end{acknowledgments}


\begin{thebibliography}{99}
\bibitem{footnote} If the aggregates are perfectly spherical,
their Coulomb barrier would have continue to grow with
their size regardless of the screening length (See Eq. (\ref{eq:vmaxrs})).
However, due to their fractal nature, the aggregate surface
where they touch each other has the same curvature as that
of an aggregates with size $r_s$. Thus, the Coulomb barrier
for two aggregates with size ${\cal R}>r_s$
remains the same as the Coulomb barrier between two aggregates
with size $r_s$ which is equal $\varepsilon r_sU^2$.
\end{thebibliography}
\end{document}